\documentclass[twoside,a4paper,twocolumn]{article}
\usepackage{fleqn,epsf,graphicx}
\usepackage{authblk}
\usepackage[utf8]{inputenc}
\usepackage[english]{babel}
\usepackage{anysize}
\usepackage{fancyhdr}
\usepackage{calc}
\usepackage{verbatim}
\usepackage{cite}
\usepackage{amsmath}
\usepackage{indentfirst}
\usepackage{balance}
\usepackage{epsfig}
\usepackage{breqn}
\usepackage{setspace}

\begin{document}


\title{\bf Time calibration of the J-PET detector}
\author{\small M.~Skurzok$^{a}$, M.~Silarski$^{a}$, D.~Alfs$^a$, P.~Bia{\l}as$^a$, Shivani$^a$,
C.~Curceanu$^b$, E.~Czerwi\'nski$^a$, K.~Dulski$^a$, A.~Gajos$^a$, B.~G{\l}owacz$^a$, M.~Gorgol$^a$,
B. C.~Hiesmayr$^c$, B.~Jasi\'nska$^d$, D.~Kisielewska-Kami{\'n}ska$^a$, G.~Korcyl$^a$, P.~Kowalski$^e$, T.~Kozik$^a$,
N.~Krawczyk$^a$, W.~Krzemie\'n$^f$, E.~Kubicz$^a$, M.~Mohammed$^{a,g}$, M.~Pawlik-Nied\'zwiecka$^a$,
S.~Nied\'zwiecki$^a$, M.~Pa\l{}ka$^a$, L.~Raczy\'nski$^e$, J.~Raj$^a$, Z.~Rudy$^a$, N.~G.~Sharma$^a$,
S.~Sharma$^a$, R. Y.~Shopa$^e$, A.~Wieczorek$^a$, W.~Wi\'slicki$^f$, B.~Zgardzi\'nska$^d$,
M.~Zieli\'nski$^a$, P.~Moskal$^a$}
\affil{$^a$ Institute of Physics, Jagiellonian University {\L}ojasiewicza 11, 30-348 Krak\'ow, Poland}
\affil{$^b$ INFN, Laboratori Nazionali di Frascati, 00044 Frascati, Italy}
\affil{$^c$ Faculty of Physics, University of Vienna, 1090 Vienna, Austria}
\affil{$^d$ Institute of Physics, Maria Curie-Sk\l{}odowska University, 20-031 Lublin, Poland}
\affil{$e$ Laboratory for Information Technologies, National Centre for Nuclear Research, 05-400 Otwock-\'Swierk, Poland}
\affil{$^f$ High Energy Physics Division, National Centre for Nuclear Research, 05-400 Otwock-\'Swierk, Poland}
\affil{$^g$Department of Physics, College of Education for Pure Sciences, University of Mosul, Mosul, Iraq}

    \maketitle
		\twocolumn[
  \begin{@twocolumnfalse}
    \begin{abstract}
The Jagiellonian Positron Emission Tomograph (J-PET) project carried out in the Institute of Physics
of the Jagiellonian University is focused on construction and tests of the first prototype of PET
scanner for medical diagnostic which allows for the simultaneous 3D imaging of the whole human body
using organic scintillators. The J-PET prototype consists of 192 scintillator strips forming three
cylindrical layers which are optimized for the detection of photons from the electron-positron
annihilation with high time- and high angular-resolutions. In this article we present time calibration and synchronization of the whole J-PET detection system by irradiating each
single detection module with a $^{22}\hspace{-0.03cm}\mbox{Na}$ source and a small detector providing
common reference time for synchronization of all the modules. 
    \end{abstract}
\end{@twocolumnfalse}
  ]
\section{Introduction}

Positron emission tomography (PET) imaging is a very important tool in medical diagnostics, in particular in oncology, cardiology, neurology, gastrology and psychiatry. Currently, all commercial PET devices are built with scintillation crystals~\cite{Slomka,Vandenberghe,Karp}. 
There are few known methods for PET scanners using time calibration. Time-of-flight~(TOF)-PET synchronisation is carried out with radioactive isotopes like sodium or germanium, placed inside the PET device, typically in 
its geometric center. The gamma quanta from radioactive source are scattered (due to applied shield) allowing synchronisation of all PET components~\cite{Griesmer,Laurence}. There are also methods for time synchronization using several radioactive sources simultaneously~\cite{Muehllehner} as well as using a rotating source along the scintillation chamber~\cite{Stearns}.  \\
J-PET is the first positron emission tomography scanner built from plastic scintillators which,
as organic detectors, are relatively cheap and easy to shape as well as are characterized by very
good time measurement resolution. In the J-PET detection system, information about the place of
$\gamma$ quanta interaction is extracted solely from timing measurement instead of energy deposition
measurement~\cite{Moskal_NuclInstr2014,Gajos_2016,Kaminska_2016,Moskal:2016moj,Moskal:2016ztv,Moskal:2015nim,Raczynski:2014nim,Raczynski:2015nim,Raczynski:2017pmb}. Therefore, it is crucial to perform precise time calibration of the detection setup.\\
The time calibration for the J-PET scanner is carried out based on measurements performed with
a reference detector and radioactive sodium source. Collected data allowed us to perform a time calibration
for each of the 192 scintillator strips (i. e. the time difference calibration of a single detection module),
synchronize between the strips in a single cylindrical layer, as well as synchronize between
the three scintillator layers. The time calibration method is briefly described in this report.

\section{J-PET calibration with a reference detector}
\begin{figure}[h!]
\begin{center}
\includegraphics[width=5.5cm,height=6.0cm]{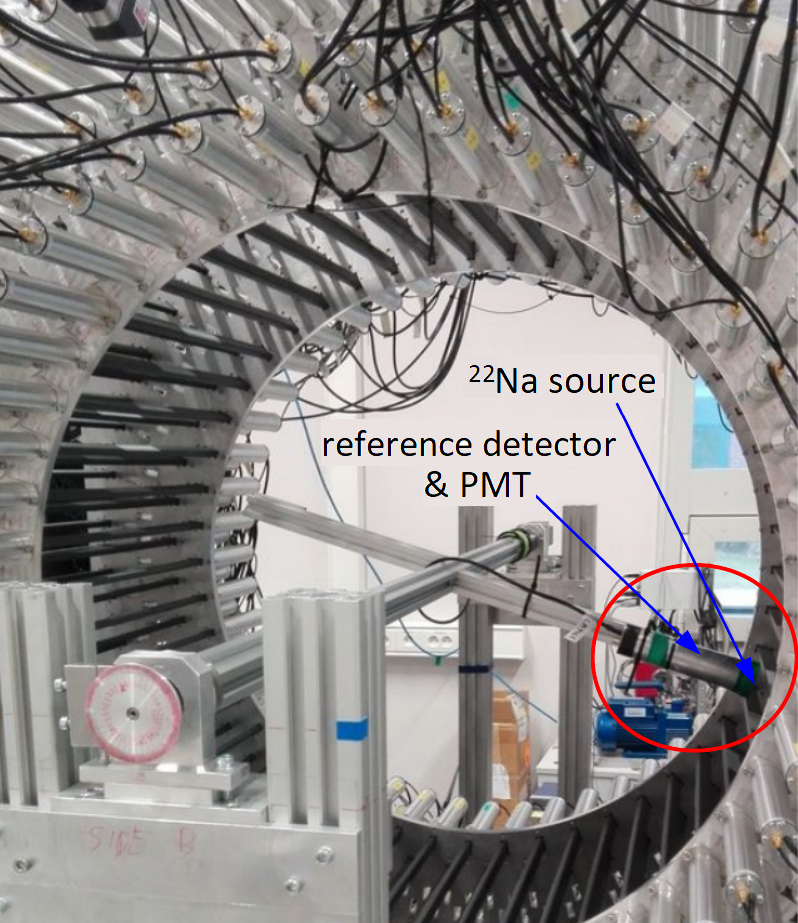}
\caption{A photo of the J-PET scanner with reference detector mounted inside the diagnostic chamber.
\label{detector_PETref}}
\end{center}
\end{figure}

\begin{figure}[h]
\begin{center}
\includegraphics[width=6.5cm,height=3.5cm]{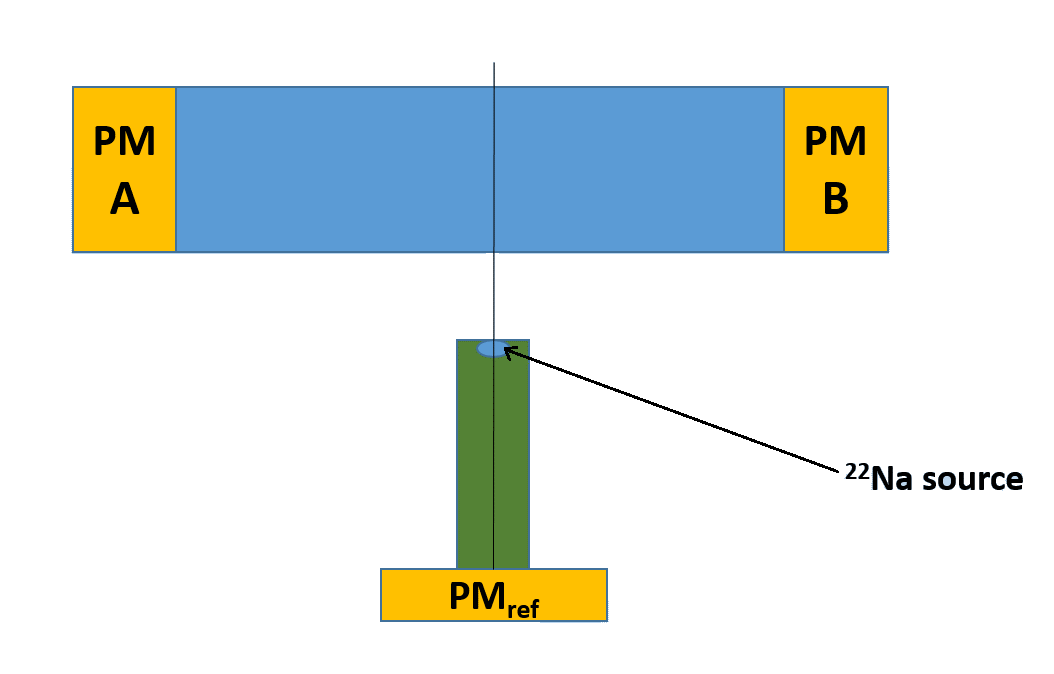}
\caption{Scheme of a measurement performed for a single J-PET detection module.
The scintillator strip, reference detector and photomultipliers are marked with blue,
green and yellow rectangles, respectively.~\label{singlestrip_meas}}
\end{center}
\end{figure}

Measurements used for the calibration and synchronization of the J-PET detector modules was performed
using a 5 $\times$ 5 $\times$ 19 mm$^3$ BC-420 plastic reference detector coupled to a single photomultiplier and with a $^{22}\hspace{-0.03cm}\mbox{Na}$ source placed on it~\cite{bednarski}. The whole system was mounted on a metal arm
inside the \mbox{J-PET} detector as shown in Fig.~\ref{detector_PETref}.
A single measurement is carried out with the reference
detector pointing at the center of the scintillator strip to be measured, schematically presented in
Fig.~\ref{singlestrip_meas}. The data for each detection module were taken in coincidence with signals
from the reference detector which, due to small size of the reference scintillator, selects a well defined
beam of gamma quanta annihilation for calibration.

\begin{figure}[h!]
\begin{center}
\includegraphics[width=7.5cm,height=5.0cm]{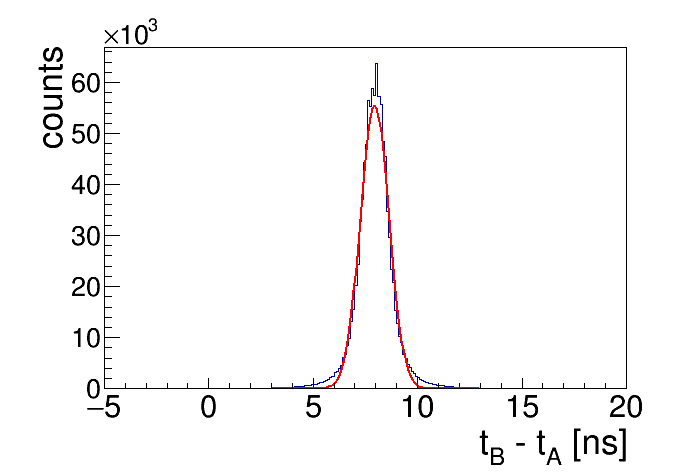}
\caption{Spectrum of the time difference between signals measured at two sides of a strip. Red curve
presents gaussian fit to the data.~\label{fit_exampleAB}}
\end{center}
\end{figure}

The measurement procedure was repeated for each of the 192 \mbox{J-PET} scintillator strips arranged
in three cylindrical layers. Since the J-PET front-end electronics are able to probe signals
at four different thresholds, and on both the leading and trailing edges~\cite{Korcyl:2016pmt,Palka:2017wms}
the calibration was done for times measured on each threshold of both signal edges.  
Collected data were analysed using the J-PET Framework software~\cite{Framework} w. r. t.
the time calibration of each separate module (so called "A-B" synchronization) and w. r. t. the TOF between scintillators.\\
%
The  calibration was performed taking advantage of the fact that the beam of selected annihilation
quanta hits each scintillator in the center. Thus, in an ideal case the difference between the
times of the signals registered at both sides of a single scintillator $\Delta t_{AB} = t_{B}-t_{A}$
must be equal zero. But in reality the measured times are shifted with respect to the true values
by some constant factors accounting for the delays in the photomultipliers and electronic components. Thus, $t_{A} = t_{A}^{true}- off_{A}$ and $t_{B} = t_{B}^{true}- off_{B}$, and the
time difference $\Delta t_{AB}$ will be non-zero. For the "A-B" synchronization ($c_{1}$)
we can determine the $\Delta t_{AB}$ distribution for each detection module. By performing a gaussian fit to each of these distributions one can extract the effective time offsets ($off$):
\begin{equation}
c_{1} = off_{A} - off_{B}.~\label{eq_1}
\end{equation}
An example of a $\Delta t_{AB}$ spectrum for one of the scintillator strips in the first layer of the J-PET detector
is presented in Fig.~\ref{fit_exampleAB}. 

In order to perform simultaneous synchronization of all modules in a single detector layer, a difference between the time of gamma quanta hit in the module and the time measured with the reference detector was determined $\Delta t_{ref} = (t_{A}+t_{B})/2-t_{ref}$. Again, a fit to the $\Delta t_{ref}$ distributions gives the common reference time for all the modules, i.e. time synchronization, and gives the calibration constant ($c_{2}$) related to the time offsets on both sides of a strip in the following way: 
\begin{equation}
c_{2} = -(off_{B} + off_{A})/2.~\label{eq_2}
\end{equation}

\begin{figure}[h!]
\begin{center}
\includegraphics[width=7.5cm,height=5.0cm]{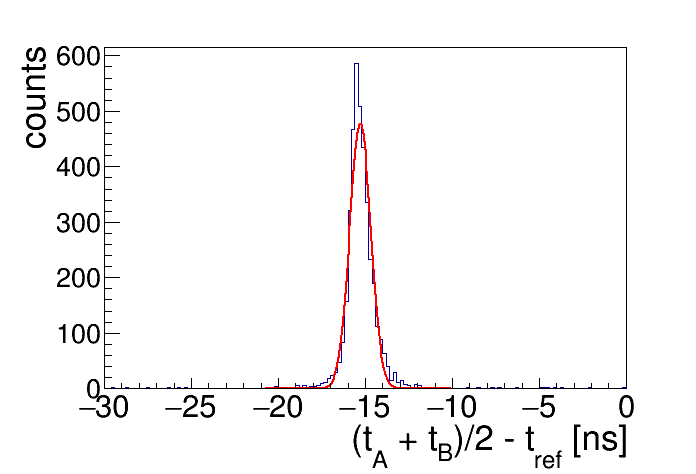}
\caption{Spectrum of time difference between qamma quanta hits in a J-PET module and reference
detector $\Delta t_{ref} = (t_{A}+t_{B})/2-t_{ref}$. Red curve presents gaussian fit to the data.
\label{fit_exampleABRef}}
\end{center}
\end{figure}

An example of a raw (without any selection conditions) $\Delta t_{ref}$ spectrum for a strip in the first J-PET layer is presented
in Fig.~\ref{fit_exampleABRef}. 

Solving the set of equations~\ref{eq_1} and~\ref{eq_2} gives finally the following time offsets: 
\begin{equation}
off_{A}=c_{1}/2 - c_{2} ~\label{eq_offsets1}
\end{equation}
\begin{equation}
off_{B}=-c_{1}/2 - c_{2}~\label{eq_offsets2}
\end{equation}
Synchronization between layers was carried out with respect to the first internal layer.
The constant $c_{2}$ (in Eq.~\ref{eq_offsets1} and~\ref{eq_offsets2}) was then corrected for strips in the other layers ($L$)
with time constants $\Delta t_{L2-L1}$ and $\Delta t_{L3-L1}$ corresponding to the time elapsed for
gamma quanta traveling from layer 1 to layer 2, or to layer 3, respectively. Time differences between
layers were calculated based on known distances between layers and were found to be equal to
$\Delta t_{L2-L1} = 0.1418 \pm 0.0033$ [ns] and
$\Delta t_{L3-L1} = 0.5003 \pm 0.0033$ [ns].  
%

\begin{figure}[h!]
\begin{center}
\includegraphics[width=7.0cm,height=4.5cm]{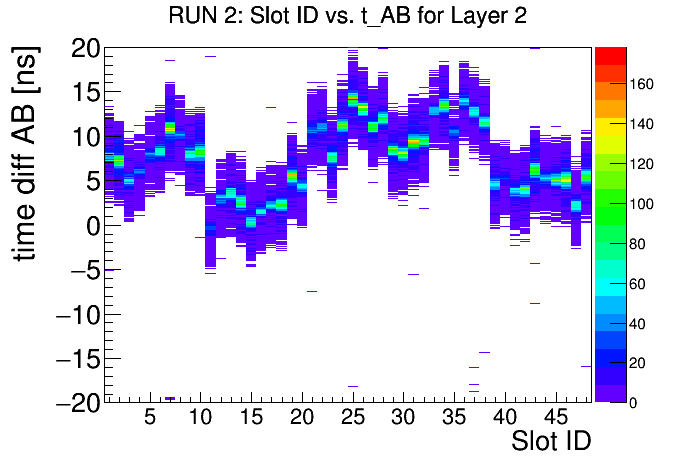}
\includegraphics[width=7.0cm,height=4.5cm]{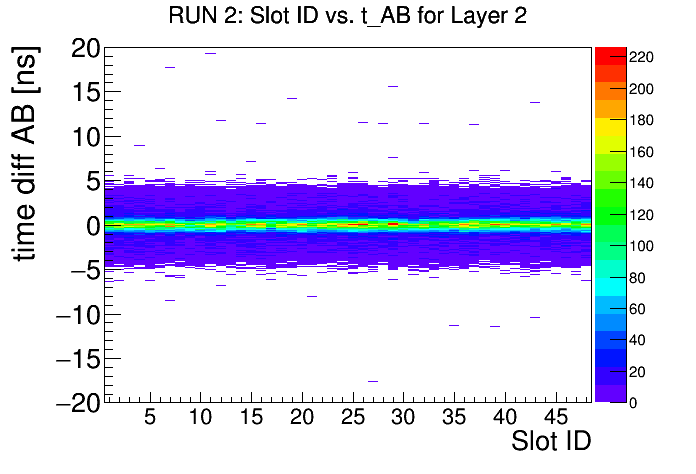}
\caption{Spectra of the $\Delta t_{AB}$ time difference as a function of the detection module number
for the \mbox{J-PET} layer 2 obtained using collimated $^{22}$Na source before (upper panel) and after
calibration (lower panel).
~\label{tAB_Run2}}
\end{center}
\end{figure}

The time calibration method was validated with independent measurements performed using a collimated
$^{22}$Na radioactive source installed in the geometrical center of the J-PET barrel~\cite{Kubicz:2016ebe}.
As an example, in Fig.~\ref{tAB_Run2} we show the $\Delta t_{AB}$ time difference for each strip of layer 2
before (upper panel) and after applying calibration constants (lower panel). As one can see, the $\Delta t_{AB}$
is distributed around zero as expected for a properly calibrated detector. 

\begin{figure}[h!]
\begin{center}
\includegraphics[width=7.0cm,height=4.5cm]{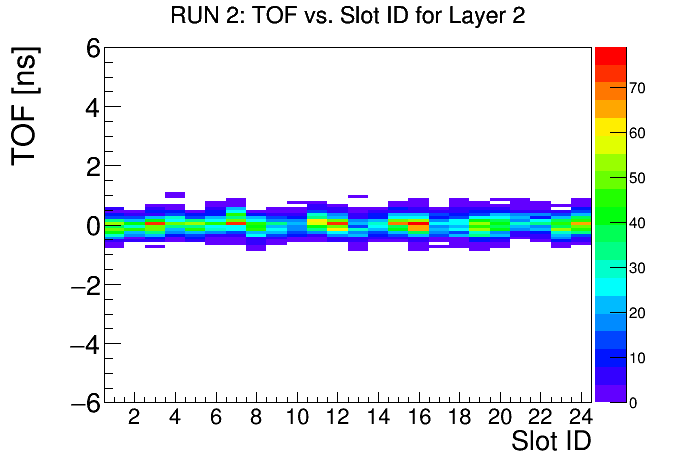}
\caption{Spectrum of TOF (time of flight) versus ID of the pair of opposite slots in the 2nd \mbox{J-PET} layer obtained
for data taken with the collimated $^{22}$Na source after the time calibration.
~\label{TOFcoinc_Run2}}
\end{center}
\end{figure}

The synchronization of modules can be
checked by studying the TOF of annihilation gamma quanta for two modules located opposite
to each other. TOF is defined as the difference between the measured times of two back-to-back gamma quanta hits.
Fig.~\ref{TOFcoinc_Run2} shows the TOF spectrum for pairs of modules in layer 2 after the synchronization.
As in the previous case we expect the distribution to be peaked around zero for all the modules since the source
was placed in the geometrical center of the detector.


\section{Summary}
We have presented the method used for time calibration of the J-PET detector. It is based on data taken by
irradiating each detector module with a radioactive sodium source in coincidence with reference detector.
This data was used to calibrate the time difference measurement within each single module and for time
synchronization of modules in all detector layers. The method was validated with independent measurements using a collimated $^{22}$Na source placed in the center of the detector, demonstrating that the developed procedure
gives satisfying results. There are other methods which may be used for the J-PET calibration and
monitoring. For example, a measurement referenced to cosmic radiation~\cite{Silarski:2013yya}; the performance and limitations of this and other methods is now under investigation.

\section{Acknowledgements}

The authors acknowledge the technical support by A. Heczko, W. Migda{\l}, the financial support from the Polish National Center for Development and Research through grant INNOTECH-K1/IN1/64/159174/NCBR/12, the EU and MSHE Grant no.~POIG.02.03.00-161 00-013/09, and the National Science Center Poland based on decision number DEC-2013/09/N/ST2/02180 and UMO-2016/21/B/ST2/01222. B. C. Hiesmayr acknowledges gratefully 
the Austrian Science Fund FWF-P26783.


\end{document}